# Naturalistic Metaphysics and the Parity Thesis: Why Scientific Realism Doesn't Lead to Realism about Metaphysics

Raoni Arroyo* [1,2,4] and Matteo Morganti [3,4]

* Corresponding author: raoniarroyo@gmail.com

[1] Centre for Logic, Epistemology and the History of Science (CLE). Campinas, Brazil.

[2] Graduate Program in Philosophy, University of Campinas (UNICAMP). Campinas, Brazil.

[3] Department of Philosophy, Communication and Performing Arts. University of Roma Tre. Rome, Italy.

[4] Research group in Logic and Foundations of Science (CNPq). Florianópolis, Brazil.

**Abstract**

In recent work, Nina Emery has defended the view that, in the context of naturalistic metaphysics, one should maintain the same epistemic attitude towards science and metaphysics. That is, naturalists who are scientific realists ought to be realists about metaphysics as well; and naturalists who are antirealists about science should also be antirealists about metaphysics. We call this the 'parity thesis'. This paper suggests that the parity thesis is widely, albeit often implicitly, accepted among naturalistically inclined philosophers, and essentially for reasons similar to Emery's. Then, reasons are provided for resisting Emery's specific inference from scientific realism to realism about metaphysics. The resulting picture is a more nuanced view of the relationship between science and metaphysics within the naturalistic setting than the one which is currently most popular.

# Introduction: the parity thesis

Let us begin by defining two key notions that will be discussed throughout this article: 'realism' and 'naturalism'. The sense of realism that will be relevant here is as the view that there are good reasons for believing that the best theories in a given domain track the truth (obviously enough, antirealism will be intended accordingly[1]). Naturalistic metaphysics is instead a meta-metaphysical stance that calls for continuity between metaphysics and science. Such a continuity can be cashed out in several ways and, although we will say more about this later, no specific characterisation is assumed or aimed for here.

Rather, our goal is to emphasise a background thesis that may underlie (and, we believe, often underlies) such a call for continuity between scientific and metaphysical theorising, especially among scientific realists, i.e., those who are realists specifically with respect to scientific theories. Having identified such a thesis and suggested reasons for considering it widely shared in the current literature, we will argue that there are reasons for resisting it. To get things started, here is a rough and ready presentation of what we called in a previous work (Arroyo & Morganti, 2025) the 'parity thesis' (PT):

> **Parity thesis (PT)**. It is more plausible for naturalistic metaphysicians to maintain the same epistemic attitude towards both science and metaphysics.

More precisely, PT is the claim that naturalism (N) establishes a biconditional connection: N → (*R(s)* ↔ *R(m)*), *R(x)* indicating that realism about *x* is regarded as justified, and 's' and 'm' being short for science and metaphysics, respectively. To avoid misunderstanding, let us make it clear at the outset that what is at stake here is not 'metaphysical realism', i.e., the view that there is an external, mind-independent reality. What PT demands is that realists about scientific theories be also realists about theories in metaphysics.[2] Importantly, the above biconditional should *not* be interpreted as strict logical implication, but rather as something weaker, such as the antecedent making the consequent significantly more plausible or compelling, albeit in an entirely defeasible manner. Obviously enough, once it is so understood, PT can be unpacked as follows:

---

[1] It is worth specifying that antirealists may take issue with either the semantic component of realism (thus rejecting the thought that the relevant claims have a well-defined content, hence truth-value) or its epistemic component (contending that, even if those claims have well-defined truth-values, we systematically lack good reasons for believing that they are true). This distinction will not play any particular role in what follows.

[2] In other words, the relevant sense of realism is as a thesis *about* metaphysics, not *in* metaphysics (on this, see, e.g., Balaguer 2025, § 1). Whether naturalism and scientific realism require metaphysical realism is a question on which we do not need to take a stand here. It is discussed by Corti (2023) — who answers in the affirmative — and Jaksland (2020) — who disagrees.

i. N → (*R(s)* → *R(m)*)

ii. N → (*R(m)* → *R(s)*)

PT, therefore, urges naturalists who are scientific realists to be realists about metaphysics as well (as per i. Above). Similarly, naturalists who are antirealists about science should also be antirealists about metaphysics (as per ii. above, via modus tollens).

To the best of our knowledge, PT was explicitly formulated and defended only recently by Emery (2023). However, we think that the general attitude it conveys may plausibly be regarded as pervasive in the philosophical community. To give at least a general idea before getting deeper into the issue, the basic thought is the following. By looking at the more or less recent literature, it seems to us, those philosophers who 'take science seriously' and consequently endorse some form of naturalism split almost entirely into two camps. Scientific realists, on the one hand, typically use naturalism to extend their epistemic stance towards the possibility of science discovering objective truths about the world to (aptly constrained) metaphysics — for examples, see Ladyman & Ross (2007) and Lewis (2016). This leads them to endorse PT, without further questioning the realist framework at all. No doubt, this attitude is also supported by the traditional conception of metaphysics as an inquiry into the fundamental structure of reality, hence as an attempt to discover fundamental truths. Scientific antirealists, on the other hand, take it to be obvious that their scepticism towards claims in science applies to claims in metaphysics as well. Some of them go on to say that, since — unlike theoretical posits in science — metaphysical hypotheses are not even useful for saving the phenomena, we should dispense with them altogether. In this case, antirealism about metaphysics amounts to the elimination of metaphysics (Carnap, 1931; van Fraassen, 1991; 2002). Others (Bueno 2021, Emery 2023) accept that metaphysical hypotheses may have some utility independently of their truth. In this case, the elimination of metaphysics is avoided by adopting a non-realist attitude towards it.[3] Such an attitude, however, is accompanied by some form of scientific antirealism. Thus, unless one eliminates metaphysics altogether, again the same epistemic attitude is selected with respect to both science and metaphysics, and PT applies.

We are happy to concede this last part of the parity thesis, i.e., ii. above. That scientific antirealists who are naturalists cannot be realists about metaphysics, as a matter of fact, should appear obvious: what would be the source of a higher confidence in something more remote from the empirical input in a context where, by definition, science is regarded as providing the benchmark for epistemic warrant? Another way of putting the same point is by means of Chakravartty's (2025, § 11.2) assessment of epistemic risk: given that discussions in metaphysics often entail — due to their greater generality — a higher degree of epistemic risk compared to the core issues underpinning scientific realism, it seems only fair to assume that those willing to embrace more epistemic risk will ipso facto accept lower degrees of it as well,

---

[3] Bueno is clearly sympathetic to empiricism in the philosophy of science, while Emery — as we will see — explicitly argues that pragmatism in metaphysics and pragmatism in science go hand in hand, but a realist application of PT is also perfectly viable.

but not the converse.[4]

The more controversial point (and our focus in this paper) is whether naturalists who are scientific realists should really regard realism about metaphysics as the only — or, at any rate, the most plausible — option, i.e., i. above. Obviously enough, if this turned out not to be the case, then PT would collapse. This, we will argue, is exactly what should happen. In order to get there, we will proceed as follows. In § 1 we provide some limited evidence in support of the claim that the following is a widespread belief: given naturalism, realism about science naturally leads to realism about metaphysics. In § 2 we present Emery's more explicit argument to the effect that scientific realism is both necessary and sufficient for realism about metaphysics. In § 3 we present reasons for resisting Emery's argument, hence PT, and in particular the idea that extending scientific realism to realism about metaphysics is a compelling, if not inevitable, move in a naturalistic context. In § 4, we close with some brief additional considerations.

## 1. Naturalism, scientific realism and the aim of metaphysics

What is the aim of metaphysical theorising? As traditionally conceived, the answer to this question is cashed out in realist terms, which Stanford puts as follows:[5] "The admirably immodest goal of metaphysical inquiry has always been to answer our deepest questions concerning the fundamental constitution, organization, and character of the world" (Stanford 2017, p. 137). In naturalistic metaphysics, such a goal might be achieved in at least three distinct ways, says Stanford:

a)      Naturalised metaphysics qua *scientific metaphysics*;

b)      Naturalised metaphysics qua *complementary metaphysics*;

c)      Naturalised metaphysics qua *metaphysics of science*.

Let us briefly analyse each of these perspectives on naturalistic metaphysics in turn, emphasising their realist commitments. Naturalised metaphysics conceived as scientific metaphysics in Stanford's sense should have its methodology and content "derived from" science. Think of Ladyman and Ross' conception of naturalised metaphysics as a *substitute* for analytic (i.e., non-naturalistic) metaphysics. Since the latter, in their diagnosis, "[…] fails to qualify as part of the enlightened pursuit of objective truth, and should be discontinued" (Ladyman & Ross, 2007, p. vii), the authors urge a re-definition of metaphysics: "By 'metaphysics' we mean something more limited and carefully constrained. We refer to the articulation of a unified world-view *derived from the details of scientific research*" (Ib., p. 65, emphasis added).

---

[4] We thank Anjan Chakravartty for pointing this out in a personal communication.
[5] A more comprehensive collection of indicative quotations can be found in McKenzie (2022, p. 1).

The key point here is that metaphysical hypotheses and claims are taken to be more or less directly derivable from science, so that — in at least some cases — sufficient knowledge of the relevant science also suffices for solving metaphysical problems (where this is not the case, obviously enough, metaphysics should be simply set aside). Stanford rightly notices that such a goal cannot be achieved without assuming scientific realism in the background:

> "Perhaps most importantly, if we wish to see our best scientific theories as giving us answers to the traditional concerns of metaphysics, then it seems we must first embrace a quite strong version of scientific realism. That is, we can only be as confident in our answers to metaphysical inquiries as we are in the truth (and completeness) of the science from which they are derived (in whatever way)." (Stanford, 2017, p. 131)

For those who subscribe to naturalistic metaphysics qua *scientific metaphysics*, then, realism about metaphysics requires realism in science. It is only insofar as science is assumed to deliver the truth about the world that one's naturalistic metaphysical lessons may be derived from it. However, since i) some form of scientific realism is assumed, ii) metaphysics is intended as the pursuit of fundamental truths and iii) metaphysical claims are regarded as directly derivable from science, it also follows that scientific realism makes realism about at least some metaphysical claims justified.[6]

Still bearing on Stanford's three-way classification, supporters of naturalised metaphysics qua *complementary metaphysics* (b) above) believe that science and metaphysics share the same *methodology*. It is plausible to think that such shared methodology essentially boils down to inference to the best explanation based on empirical and extra-empirical factors. That extra-empirical, or 'theoretical' virtues play a key role is, for instance, contended by Paul, who explicitly states that both science and metaphysics make use of "[…] *a priori* reasoning involving simplicity, elegance and explanatory strength" (Paul, 2012, p. 19). According to Paul, the same is not true for the *content* of the two disciplines, as the questions addressed by metaphysics are distinct from those addressed in scientific inquiry. While Paul grants that "[b]oth fields are interested in discovering truths about entities or features of the world" (Ib., p. 9), she acknowledges that metaphysics "involves features of the world that are metaphysically prior to those of the scientific account." (Ib., p. 5). Hence, science and metaphysics are *complementary* in the sense that they share essentially the same methodology and formulate conjectures of different depths about the same object of study, i.e., reality.

On this account, epistemic value is not *transferred* from science to metaphysics. The two disciplines just happen to be, as Saatsi puts it (without accepting the view), of a piece: "[…] metaphysics, like theoretical science, is just further (albeit more abstract or general) theorising about the unobservable world" (Saatsi, 2017, p. 167). Emery eloquently puts it as follows. First, she suggests that most contemporary metaphysicians are 'content naturalists', meaning that they accept the maxim that one "[…] should not accept metaphysical theories that conflict with the content of our best scientific theories" (Emery, 2023, p. 10). In particular, content naturalists believe that:

> "[…since] science and metaphysics are about the same domain (what the world is like) and science has proved itself highly successful in answering questions about that domain, then metaphysicians should not take on commitments that conflict with the content of our best scientific theories." (Emery, 2023, p. 46)

---

[6] This is no doubt a *radical* naturalistic metaphysics (see Andersen & Arenhart, 2016; Morganti & Tahko, 2017; Guay & Pradeu, 2020). The obvious objection to it is that — regardless of the issue of realism — metaphysical lessons can hardly be extracted *directly* from scientific theories.

Next, Emery argues that if one is a content naturalist then one should also be a methodological naturalist, i.e., believe that metaphysicians should employ the same methodology that scientists use — thus endorsing naturalism as complementary metaphysics in Stanford's sense. The reason for this is basically that if one doesn't consider the methodology of science the best guide to formulating conjectures about reality, then one doesn't have reasons for being a content naturalist in the first place. According to Emery, however, this implies that one's epistemic attitude towards science can, and should, be extended to metaphysics. On the basis of this, Emery concludes that scientific realists who embrace naturalism should be realists about metaphysics.[7]

In Stanford's third and final conception, naturalistic metaphysics qua *metaphysics of science* further articulates the metaphysical content scientific theories *already have, at least partially and/or potentially*. It is a proposal that, as Stanford (2017, p. 137) puts it, "applies metaphysics to science itself". It is the kind of metaphysics one does, for instance, when trying to find out whether or not quantum entities are individual objects by applying the tools and theories developed in metaphysical theorising (cf. French & Krause, 2006). In contrast with scientific metaphysics, at least on our understanding of Stanford's reconstruction, instead of *extracting* the metaphysics out of the scientific discourse, as it were, on this third conception of naturalistic metaphysics one *adds* metaphysics to science. For example, although quantum mechanics certainly has a metaphysical component to it insofar as it concerns the behaviour of entities with properties located in space and time, nothing in the theory tells us in a conclusive way whether, say, quantum objects are individuals or not. What the metaphysician of science does is, then, to apply to it specific views on objects and identity which were independently developed by philosophers, possibly providing reasons for preferring one over the others in the quantum domain. On such a naturalistic approach to metaphysics, therefore, metaphysical theorising — qua metaphysics of science — is a metaphysical addition to what science tells us, where science guides metaphysics and puts significant constraints on metaphysics without entailing it.

---

[7] As a possible objection to Emery's line of reasoning, one could argue that science provides the best methodology for formulating conjectures about reality, while maintaining antirealism both towards science and metaphysics. However, Emery's point is that, *if* one believes that science provides some access to reality then, due to methodological continuity, there's no plausible reason to hold an antirealist stance about metaphysics. Indeed, as we will see later, Emery does acknowledge antirealism about both science and metaphysics — without elimination — as an option. We thank an anonymous referee for raising this point.

At a first glance, one could think that there's no realist commitment here (neither with respect to science nor to metaphysics). As Stanford writes, one may contend that the metaphysics of science deals with "questions about theories and not about the world" (Stanford 2017, p. 137). What is at issue, that is, is (to stick to the above example) whether or not quantum objects as described by the theory are to be cashed out in terms of individuality, not whether the world is correctly represented in such a way. After all, quantum mechanics might be replaced by some other theory in the future. However, a realist commitment to metaphysical posits seems to be regarded as obvious by many authors even on this third conception of naturalised metaphysics, at least as long as scientific realism is endorsed. For instance, McKenzie — who is arguably a supporter of naturalistic metaphysics in the sense of 3) above (see French & McKenzie, 2012, 2015) — says:

> "[…] since the present target of investigation is whether metaphysics can be thought of as making progress on the assumption that science is, we will assume that science makes progress by producing better approximations to the truth. Nor should this characterization be regarded as at all controversial; the production of better approximations to the truth is standardly presented as the core commitment of scientific realism." (McKenzie 2020, p. 12, original emphasis)

Clearly, if progress in science is to be intended in the realist sense, and metaphysics makes progress to the extent that science does (by articulating the metaphysical aspects of the latter), it follows that progress must be intended in the realist sense in metaphysics too. To be clear, in her paper McKenzie goes on to raise objections against this generalised realist assumption — which, consequently, she does not share. The point, however, is that McKenzie's argument is presented exactly as going against an attitude that she regards as so widespread that it is typically considered unproblematic.

Indeed, one may go as far as to claim that a literature survey shows that most naturalistically oriented authors who consider it worthwhile to add a metaphysical gloss to science do so because they assume scientific realism and, based on this, regard realism about metaphysics as an obvious choice.[8] For instance, French says:

---

[8] In her paper, McKenzie ends up arguing that we are not justified in being realist about metaphysics, even in its science-based variety. However, she does this on the basis of the incompleteness of current science, which is arguably a good reason for rejecting scientific realism as well — which is compatible with PT. At any rate, here we are not discussing McKenzie's specific views. Rather, we are noting that

> "[…] metaphysical minimalism raises concerns as to whether we obtain the clear understanding of how the world is that we associate with scientific realism. […T]he realist cannot rest content with epistemology but must seek an understanding articulated in metaphysical terms." (French, 2014, p. 7)

French champions 'deep realism', viz., the idea that the addition of genuinely metaphysical claims — say, about the nature or properties or the (non-)individuality of microscopic physical systems — is a necessary condition for one to hold a truly realist stance in science — i.e., one that does not collapse onto a 'shallow realism' which is, in fact, a 'closet empiricism' (French, 2014, p. 48). To this purpose, he claims[9] that the scientific realist ought to use metaphysics as a toolbox, providing the instruments for the definition of a precise ontology underpinning one's favourite scientific theory. While French endorses a peculiar view of science and metaphysics, and one that is by no means widely agreed upon, his basic intuition about the fact that scientific realism does require some amount of realism about metaphysics seems instead quite popular.

Ney's (2012) 'neo-positivist' metaphysics, for instance, goes in the same direction: she explicitly states that "[o]ur metaphysical project depends on the attitude in general of the physics community being realist" (Ney 2012, p. 65). Thus, whatever metaphysical gloss we add to scientific theories, this is done by scientific realists on the assumption that the latter are at least partially true descriptions of reality. If this were not the case, the idea seems to be, there would be no point in adding anything to science in the first place. At the same time, crucially, the very fact that one is a scientific realist and a viable methodology for naturalistic metaphysics has been identified makes realism towards this latter addition justified. As a matter of fact, that the metaphysical project is said to 'depend' on scientific realism seems to be because realism about metaphysics is regarded as a crucial desideratum, and one that naturalistic methodology can satisfy.

In view of the above, we conclude that also on the third and last conception of naturalised metaphysics identified by Stanford, typically, scientific realism is straightforwardly merged with realism about metaphysics.

More generally, it seems to us that not only do the above examples and considerations indicate that naturalists regard scientific realism as a *necessary* condition for realism about metaphysics — no doubt a rather popular, and uncontroversial, claim that may be regarded as one of the characterising features of naturalism itself (Ladyman, 2012, p. 34; Bryant, 2020a, p. 1868; 2020b, p. 29;[10] Jaksland, 2023a, p. 6). What we regard as an equally widespread — albeit

---

she uses as a (polemic) starting point for her argument the assumption that naturalism is a good bet for those who want to be realists in both the scientific and the metaphysical domain.

[9] In the abovementioned joint work with McKenzie, see French & McKenzie (2012; 2015).

[10] Bryant (2024) claims that scientific realism is not necessary for naturalistic metaphysics. But, of course, this is compatible with PT as long as a naturalistic approach to metaphysics does not entail realism about it.

certainly less explicit — attitude, at least among scientific realists who subscribe to naturalism, is one that can be expressed as follows: under the right conditions, one's realism about scientific theories and posits extends to, and in turn benefits from, realism about metaphysical theories and posits. In this sense, scientific realism is also *sufficient* for regarding realism about (some, aptly grounded in science) metaphysical hypotheses as justified. The above quotation from French is particularly telling in this sense. For, notice, it *starts* from scientific realism and quickly ends up endorsing realism about metaphysics as a way to augment the 'toolbox' at our disposal for inquiring into the fundamental structure of reality. It is clear that French intends this in a modal/normative sense: he claims that scientific realists "cannot rest content with epistemology" and "must seek" a metaphysical understanding. But this unquestionably amounts to the idea that endorsing scientific realism is sufficient for realism about metaphysics.

Further textual evidence is not hard to find. Saatsi, for instance, argues that, for the traditional scientific realist — with whom he disagrees — the various interpretations of quantum mechanics are "metaphysical alternatives" to what "fundamental reality" might be, "and for all we know one of them might depict the world more or less correctly" (Saatsi, 2019, pp. 145–146). This is in line with Chakravartty's account of what scientific realism is:

> "[…] is the scientific realist a metaphysician? Here we come to perhaps the deepest matter of all. Ultimately I think that the answer to this question is 'yes' […] some scientific realist beliefs are surely, if properly called 'metaphysical,' metaphysical in a pre-Kantian way, in that they pertain (in intention, at least) to the world itself, the noumenal world, not merely the world as we fashion it through human ways of knowing" (Chakravartty & van Fraassen, 2018, p. 24).

Chakravartty also states that

> "[…] many contemporary philosophers of science and, prominently among them, many scientific realists do advocate beliefs concerning things that philosophers today would still regard as metaphysical, including beliefs about properties, causation, laws of nature, de re modality, and so on. Indeed, philosophical defences of the reasonableness of believing in the sorts of scientific entities and processes that are not generally considered metaphysical today, such as genes and gene transcription, often make recourse to views about things that are regarded as falling under the purview of metaphysics, such as causation, modality, and so on" (Chakravartty, 2013, p. 28, original emphasis).

And a 'generalised' sort of realism, about both science and metaphysics, also seems to underpin more focused work such as, for instance, recent attempts to merge many-worlds quantum theory and modal realism (Wilson 2020; for a general overview on modal issue at the boundary between science and philosophy, see Bryant & Wilson 2024).

If we are right, and PT is shared by the great majority — if not the totality — of scientific realists who endorse naturalism, this does not come as a surprise. After all, naturalised metaphysics doesn't change what is traditionally conceived as the goal of metaphysics, viz. the search for fundamental truths about reality. To the contrary, it changes the methodology of how one is supposed to achieve such a goal, from mostly or purely a priori methods to empirically informed, science-based ones. Indeed, as Ladyman states, naturalistic metaphysics aims for a "reform, not abolition" of metaphysics (Ladyman, 2017, p. 142). It is therefore plausible to think that metaphysical naturalists who are non-eliminativists about metaphysics agree with this common characterisation. And that, consequently, they regard naturalism as a way to make it practicable in a broadly empiricist setting, where a posteriori knowledge is given a privileged position. What this entails is, crucially, that — while of course naturalism by itself is compatible, as we have seen, even with eliminative stances towards metaphysics — as soon as naturalists endorse scientific realism, they are naturally led towards realism about metaphysics.

In view of the foregoing, we believe that there are good reasons for thinking that PT is typically endorsed by naturalists. In what follows, however, we will focus mainly, if not exclusively, on the only explicit argument in favour of PT that — to our knowledge — can be found in the literature, due to Emery (2023). This will enable us to achieve at least a minimal, uncontroversial goal. Those who have not been persuaded by the above arguments about naturalism and realism, in particular, can read the sections to come as specifically devoted to a critical analysis of Emery's perspective on PT. For, all the other readers, the intended target will instead be a more general attitude towards science and metaphysics. Regardless of this, we believe the following discussion of PT is relevant beyond the assessment of Emery's work, as it affects the range of naturalistic views that one may find viable.

Before moving on, however, let us add a few remarks and clarifications. To begin with, it is an important question whether realism about metaphysics is well motivated in itself. For instance, while Ladyman & Ross (2007; see also Ladyman, 2017) and Ney (2012; 2020) are optimistic, Arenhart & Arroyo (2021; see also Arroyo & Arenhart, 2024) argue that the prospects of achieving truth in metaphysics are grim due to metaphysical underdetermination. A similar thesis is endorsed by Jaksland (2023b). Others still, are in principle open to naturalistic metaphysics, yet "think there is a real possibility that the activity that we call 'metaphysics' should turn out not to constitute a viable form of inquiry at all, either empirical or non-empirical" (Melnyk 2013, p. 81).[11] This will be important in what follows but does not directly affect what we said about PT which — notice — is a biconditional that is entirely neutral with respect to the truth or falsity of its components.

---

[11] Notice that even Melnyk's pessimism does not necessarily represent a counterexample to our claim that PT is a generally accepted thesis. Besides the fact that we didn't suggest that PT is *universally* accepted and will in any case focus on Emery's specific endorsement of it in what follows, it should first be established whether Melnyk is a scientific realist. In any case, Melnyk's reasons for pessimism have to do with underdetermination and the in principle impossibility for genuinely metaphysical questions to receive determinate answers. This indicates that he agrees with the idea that realism and elimination are the only viable options vis à vis metaphysics — which is something we will take issue with later in the paper.

Another important issue is what exactly qualifies as metaphysics. Ney (2012, p. 63) argues, for instance, that a commitment to the truth of Lorentz invariance or the Born Rule should count as a naturalistic metaphysical commitment based on indispensability considerations related to the practice of physics. On the other hand, one may follow those who draw a distinction between ontology and metaphysics (see, again, Arenhart & Arroyo, 2021; Arroyo & Arenhart, 2024), and restrict the latter to distinctively philosophical questions such as those surrounding, e.g., universals, possible worlds, identity, characterised by the use of a sui generis vocabulary and extra-scientific notions and categories, and not limited to questions concerning what exists. In what follows, we will assume the latter characterisation of metaphysics, if only because it has the welcome consequence that there is no overlap between the ontological commitments that are suggested by science 'directly', in its own language (be it to Lorentz invariance, the Born Rule, electrons, dinosaurs or viruses) and those that metaphysics adds to them (for example, to primitive identities for material objects, perdurantism about persisting objects or essentialist natural kinds). Such an overlap may render PT trivial, at least in specific instances. Indeed, those who prefer the former conception of metaphysics, and consequently regard commitment to the truth of the Born Rule or the existence of electrons as genuinely metaphysical, are invited to restrict the considerations to follow to non-trivial versions of PT anyway. That is, versions in which the truth of PT does not simply follow from the fact that, say, scientific realism about electrons is warranted if and only if metaphysical realism about electrons is.

Finally, let us parry another potential objection: we are not arguing that those who endorse a) the traditional conception of metaphysics as the search for fundamental truths share the opinion that b) the naturalisation of metaphysics yields that, at the meta-level, only a realist epistemic attitude towards it is available. This would be trivial, as naturalism would not play any actual role — the truth of the consequent in b) would already be established by a). The point is, rather, that scientific realists typically find naturalism sufficient for extending to metaphysics the epistemic attitude they have towards science. That is, not only do they agree that metaphysics is to be intended as the search for fundamental truths. They also believe that this conception of metaphysics is supported by naturalism and that naturalisation makes it possible for metaphysics to yield actual results, i.e., there are at least some genuinely metaphysical claims supported by good reasons — viz., scientific reasons.

## 2. Emery's argument for PT

In the course of her discussion of science, metaphysics and epistemic attitudes, Emery (2023) contrasts realism with pragmatism, which she defines as follows:

> "*Pragmatism about science.* Our best scientific theories are theories that are useful for creatures like us in navigating the world. […] *Pragmatism about metaphysics.* The aim of metaphysics is to put forward theories about what the world is like that are merely useful for creatures like us." (Emery, 2023, pp. 54–55, original emphasis)

The term 'pragmatism' as it used by Emery has no direct connection with pragmatist philosophy, and is basically a synonym of instrumentalism. Since we find this latter term less ambiguous, we will use it in what follows. Now, Emery argues that it is perfectly possible for a naturalist to endorse an instrumentalist attitude with respect to scientific theories, provided that one's instrumentalism towards science *is extended to metaphysics*. In this case, *its traditional characterisation notwithstanding*, metaphysics is not regarded as a means for discovering objective truths about the world but, rather, to gain understanding, define unified models, deepen our explanations and so on. Views of metaphysics along these lines have already been formulated (see Godfrey-Smith, 2006; Paul, 2012; Rosen, 2020; Bueno, 2021; Ritchie, 2022; Emery, 2023; McSweeney, 2023; Bryant, 2024; Arroyo & Morganti, 2025).[12] What is important for present purposes is that Emery's argument in favour of this generalised instrumentalist attitude is grounded in her more general endorsement of PT. Based on her definition of 'content naturalism' as the thesis according to which the content of our best metaphysical theories should not conflict with the content of our best scientific theories, she puts it as follows:

> "If there is a mismatch between one's view about science and one's view about metaphysics — if, for instance, one is a pragmatist about metaphysics but a realist about science, or if one is a realist about metaphysics but a pragmatist about science — then it would be odd to be a content naturalist. After all, someone who endorses this kind of mismatch thinks that science and metaphysics have significantly different goals — so why should they care if their scientific claims and their metaphysical claims conflict? But if one is either a realist about both science and metaphysics, or a pragmatist about both science and metaphysics, then the simple case that I gave above for being a content naturalist still applies. So my view is: no — content naturalism doesn't presuppose realism about science. What it does presuppose is some *congruence between how one thinks about the goals of science and how one*

---

[12] The claim that one can be a pragmatist/instrumentalist about metaphysics is of course bold and frequently met with resistance. After all, it changes the goal of metaphysics as traditionally conceived, viz., as fundamental truth-seeking. Such an "instrumentalist metaphysics" wouldn't qualify as metaphysics properly for, e.g., Ladyman & Ross (2007) and Jaksland (2023b). Consequently, such authors would endorse the view that antirealism about metaphysics means eliminativism about metaphysics. Our claim in what follows will be that there is in fact defensible middle ground between realism and eliminativist antirealism about metaphysics, and that such an intermediate perspective meshes well with scientific realism. We thank an anonymous referee for raising our attention to this point, which is directly relevant for the thesis that we will put forward in the last part of the paper.

*thinks about the goals of metaphysics.*" (Emery, 2023, p. 55, emphasis added)

"Content naturalism doesn't require that you think scientists believe their theories, but it does require some congruence between the epistemic attitude that you think scientists have toward their best theories and the epistemic attitude that you think metaphysicians have toward their best theories. If you thought that scientists only *provisionally accepted* their best theories, but metaphysicians *believed* their best theories, it wouldn't make sense to be a content naturalist." (Emery, 2023, fn. 19, pp. 55–56, original emphasis)

The last part of the second quotation is rather telling. Emery proceeds with a specific case in mind, namely, a person who endorses antirealism in science but realism about metaphysics. As we have already pointed out, this would certainly be an ill-motivated move for a naturalist to make. After all, endorsing realism about metaphysics with no foundation in science for one's metaphysical hypotheses would entail that the desired *continuity* between the two enterprises is lacking, and consequently one is in fact not working in a naturalistic setting. This was put in crystal-clear terms by Hawley:

"[I]t should come as no surprise that anyone who is sceptical about the ability of science to give us knowledge of quarks and quasars will be sceptical about whether science can give us knowledge of universals and possible worlds." (Hawley, 2006, p. 454)[13]

Crucially, however, this doesn't exhaust the possible combinations of the relation between realism and antirealism in science and in metaphysics. How about *a scientific realist who is a naturalist and a non-eliminativist with respect to metaphysics but doesn't subscribe to realism about metaphysics*? Notice that this is different from naturalism coupled with scientific realism but with no metaphysics at all. This view, which can be regarded as the realist counterpart of van Fraassen's empiricism (see, e.g., van Fraassen 2002) can be plausibly attributed, for instance, to Psillos (see, e.g., Psillos, 1999, p. 165, and possibly to Melnyk, 2013). The underexplored combination of naturalism about metaphysics, scientific realism and metaphysical antirealism *without elimination* is precisely the view that we will argue for, after a critical assessment of Emery's argument for PT — which we now turn to.

---

[13] As a matter of fact, one may object based on counterexamples. For instance, one may have reasons for believing in universals based on mathematical practice and mathematical induction, while being sceptical with respect to physical unobservables due to, say, underdetermination. While this combination of realism about metaphysics plus antirealism about science is possible, however, we think it is in fact not available to the naturalist. Hardly any naturalist would agree that going from an abstract science to realism about universals is more plausible than going from a natural science to realism about a natural kind. We thank an anonymous referee for raising this objection.

Emery's starting point is the thought that most philosophers subscribe (either implicitly or explicitly) to content naturalism in the sense defined a moment ago; and that the latter entails what she calls 'methodological naturalism', i.e., the thesis according to which metaphysics and science use the same methodology. If methodological naturalism is correct, in particular, both science and metaphysics rely on empirical data and extra-empirical factors. Based on this, Emery infers that science and metaphysics share the same goals and consequently demand the same epistemic attitude. Conversely, if our epistemic attitude towards science and metaphysics were not the same, says Emery, then there would be no reason for us to endorse methodological naturalism in the first place. By modus tollens, we should consequently abandon content naturalism as well — and, with it, the whole idea that there should be a significant continuity between the two enterprises, i.e., naturalism itself. The upshot is that naturalism requires the acceptance of PT. However, we think that Emery's argument, even if valid, is not sound.

Before seeing why, one important remark is in order. One may worry that Emery's particular views on content naturalism, methodological naturalism and their implications are not shared by all naturalists. Consequently, one may also worry that Emery's argument for PT is not representative, as naturalists may endorse PT for different reasons (or even not endorse it at all). While we agree that there may be different ways of being naturalists about metaphysics and different reasons for choosing a realist (or anti-realist) epistemic attitude towards science and/or metaphysics, we nonetheless think that a number of plausible considerations can be made that are sufficient for assuaging the worry. First, whatever one's reasons for endorsing it and on what element one puts one's emphasis, naturalism necessarily entails a thesis about content (metaphysical ontological commitments cannot conflict with the ontology of science) and a thesis about methodology (metaphysics cannot proceed entirely a priori, and in particular entirely independently of the findings of science) — these two elements being jointly exhaustive. Secondly, regardless of her specific views about content naturalism and methodological naturalism, Emery argues that — however intended — naturalism in any case entails PT. Third, independently of the details of her argument, Emery's reasons for this claim have essentially to do — as we will see in a moment — with the widespread intuition that we discussed earlier: namely, that naturalism makes it plausible (in fact, for Emery, necessary) to regard science and metaphysics as two disciplines that cooperate towards the same goal and consequently warrant (or require) the same epistemic attitude. In view of this, in spite of the fact that Emery's argument for PT only expresses a specific take on naturalism and realism, we regard it as quite indicative of a much more general attitude, and consequently worth discussing in detail.

## 3. A critical assessment of Emery's argument for PT

First of all, let us reformulate Emery's argument in favour of PT in a more formal way. Let us fix the acronyms and definitions involved in the argument. N: Naturalism, i.e., the view that reality is best understood through natural scientific methods and principles. NC: content naturalism, i.e., the thesis according to which the content of our best metaphysical theories should not conflict with the content of our best scientific theories. NM: methodological naturalism, i.e., the thesis according to which metaphysics and science use the same methodology. SG: The claim that metaphysics and science share the same fundamental goals. PT: The claim that metaphysics and science require the same epistemic attitude. Based on this, Emery's argument can be formulated as follws:

1) $N \rightarrow (NC \lor NM)$

2) $NC \rightarrow NM$

3) $NM \rightarrow SG$

4) $SG \rightarrow PT$

∴ $N \rightarrow PT$

Recall, first of all, that PT is the thesis that ($N$ makes it plausible that ($R(s) \leftrightarrow R(m)$)). Emery argues, therefore, that if one is a naturalist then it is plausible to think that one should be a scientific realist if and only if one is also a realist about metaphysics.

Now, we take 1) above to be a fair characterisation of naturalism. It merely states that, in being a naturalistic metaphysician (someone, that is, who endorses N, and consequently thinks that there's some kind of continuity between science and metaphysics), one should cash out such a naturalist attitude as either accepting that our metaphysical theories should not be in conflict with our best scientific theories (NC) or that metaphysics should follow our best science in its methodology, or both. As we suggested a moment ago, while different conceptions of naturalism are possible, all of them minimally entail a thesis about ontological commitment (no conflict with science) and one about methodology (metaphysics should take into account the empirical input, hence science, as much as possible). Recall, Emery's own position is a case of naturalised metaphysics qua complementary metaphysics, but any of the three characterizations of N given in § 1 (viz., qua scientific metaphysics or qua metaphysics of science) is aptly characterised by the disjunction in step 1).

As for 2), Emery (2023, § 1.3) discusses the link between content and methodological naturalism at length and, although we believe that there is room for disagreement with her conclusions (viz., PT), we will take them for granted here. Consequently, we will assume that content naturalism does entail methodological naturalism (and that most, if not all, contemporary naturalists endorse both). Moving to 3), assuming that sameness of goals does not trivially include the sameness of the requested epistemic attitude which is supposed to come at the end of the argument, we believe it can be safely accepted too. After all, if two activities use the same methodology to cooperate in the definition of a consistent set of ontological commitments (as per content naturalism), it seems fair to regard them as sharing a common goal. What we object to is the idea that sharing a goal (as well as a methodology and a set of ontological commitments) is sufficient for granting sameness in epistemic attitude, i.e., 4) above. Before putting forward our objection to Emery's argument for PT, a few comments are in order. First of all, our argument against PT should not be intended as an argument that, via modus tollens, goes against NC or NM, hence ultimately against naturalism (i.e., the antecedent of premise 1)). Rather, we question the truth of premise 4), hence the idea that given naturalism, commitment to scientific realism rationally supports or a corresponding commitment to realism about metaphysics.

Let us then look at premise 4) in more detail. As suggested, Emery's idea is that sharing a common ontology and a general methodology entails sharing a common goal — in this case, that of uncovering the structure of reality (for realists) or achieving understanding or some other non-truth-involving aim (for instrumentalists). However, starting from shared ontological commitment, the fact that science and metaphysics work on the same set of entities, processes etc., and the latter should never require ontological commitments in conflict with those demanded by the former, by no means entails that each item in the set deserves, or demands, the same epistemic attitude. Already within the domain of the sciences, the degree of confidence in the reality of the relevant posits varies significantly: compare, for instance, electrons with gravitons or strings (still purely hypothetical), viruses with black holes, or chemical/biological kinds with, say, economic inflation. A fortiori, it is plausible to think that metaphysical posits are systematically more conjectural than scientific posits (hence, with a higher degree of epistemic risk in the sense of Chakravartty, 2025).

Something similar holds if one makes the notion of goal more general, and not strictly limited to the definition of a coherent set of ontological commitments: granted that metaphysics and science have the common aim, say, of uncovering the structure of reality, it is perfectly possible — indeed, we think, likely — that, due to the higher level of generality of the former, even in a naturalistic context the results that it achieves in the process are systematically less trustworthy than those of the latter. If the aim is, instead, purely instrumental, a similar gap is likely: for instance, science is useful for pursuing technological goals while metaphysics only for gaining understanding; alternatively, both disciplines may be considered instrumental to gaining understanding, but science is arguably able to provide a different kind of understanding — one that is more constrained, more closely linked to empirical data and testability, less likely to be disconnected from truth, and so on.

Lastly, and most importantly, Emery's thought could be that two disciplines that share the same methodology and the same object of study require the same epistemic attitude just because of that — i.e., because sameness of goals is not simply entailed by shared ontological commitment and methodology but is rather defined as the sum of these two things. However, this would warrant PT only with the help of additional assumptions — per se, mere cooperation and avoidance of conflict are hardly enough for sameness of epistemic attitude. However, consider realism: on the proposed understanding of shared goals, it looks as though realism about metaphysics would be reasonable only if the best extant arguments in favour of scientific realism were directly based on methodological considerations.[14] But this is definitely *not* the case: scientific realists and scientific antirealists clearly agree on the nature of scientific methodology, and on the fact that it is the best means to seek answers to certain questions. Disagreement remains, however, about the epistemic status of scientific claims going beyond the phenomena. Obviously enough, empirical evidence is insufficient for picking one theory over another, let alone to licence realism about any theory. As for theoretical virtues, since at least the seminal work by Kuhn (1977), they have been widely regarded as not truth-conducive.[15] It is, therefore, not surprising that a methodology based on empirical facts plus theoretical virtues does not by itself suffice to settle questions concerning the status of the relevant unobservable entities. By the same token, it comes as no surprise that those that are unanimously recognised nowadays as the strongest extant arguments in favour of scientific realism are not based on considerations concerning the methodology employed by scientists to get to the formulation of the relevant theories. In particular, they do not have recourse to theoretical virtues, and instead call into play features — such as, most notably, the ability to yield novel predictions (Alai, 2021), but also the unavailability of alternatives, the manipulability of certain entities or the 'consilience' of independent pieces of inductive reasoning — that do not have to do (if not rather indirectly) with scientific methodology, as witnessed by the fact that not all scientific theories and claims exhibit such properties — at least not in the same way. What is particularly important for us is that none of these characteristics is present when it comes to typical metaphysical hypotheses: they do not lead to novel empirical predictions, no genuinely metaphysical posit is manipulable, and so on.[16]

To be clear, we are not denying that extra-empirical factors must be taken into account in metaphysical and scientific theory choice, as underdetermination is most often unavoidable. What we are claiming is that certain additional elements that (may!) tip the balance in favour of realism in the scientific case, which are in any case distinct from theoretical virtues, appear to be lacking in the typical metaphysical case. Absent those elements, metaphysical hypotheses can be regarded as missing the kind of support required for a justified realist attitude towards them. It looks as though an analogous argument could be formulated in an antirealist/instrumentalist context — i.e., one to the effect that, shared goals notwithstanding, the instrumental import of metaphysics is significantly different from that of science. Regardless of this, at any rate, the considerations just made are sufficient for questioning the key component of PT, i.e., the move from scientific realism to the plausibility of realism about metaphysics (component i. in the introduction). From which it follows that — contra Emery and other supporters of PT — naturalism about metaphysics does *not* bring with itself sameness of epistemic attitude with respect to science and metaphysics.[17]

---

[14] We ignore the ontological/content component here because, as it should appear clear, it cannot license realism per se. Rather, it is realism about it that is at issue, and whether or not such realism is justified depends either on the other component, i.e., on the methodology of science, or — as we believe — on something else.

[15] Which is not to suggest that the issue is no longer open to debate. Dawid (2013) and Schindler (2018), for instance, provide reasons for thinking that theoretical virtues may lend at least limited support to realism about certain hypotheses. For present purposes, however, the fact suffices that scientific realism is typically not based on considerations of simplicity, parsimony and the like. For attempts to systematise theoretical virtues beyond the canonical pessimistic perspective, see Thagard (1978), Douglas (2013), McMullin (2014), Mackonis (2013) and Keas (2018).

[16] Alai (2023, p. 392) argues that the "no miracle argument from novel predictions ($NMA^{NP}$)" is the "ultimate" argument for scientific realism and that it also supports metaphysical realism (Ib., p. 377). This could be interpreted as $NMA^{NP}$ supporting metaphysical claims, and in particular the thought that there is an objective, independent external reality, hence undermining our point. This is not the case, however: first, Alai essentially argues that metaphysical realism is part of the best explanation of the success of our scientific practice, not that there are novel predictions in metaphysics; secondly, it seems uncontroversial that Alai's argument cannot be extended to other theses in metaphysics. In any case, for those who are not convinced, it is sufficient to stress that the claim being made in the text concerns *typical* metaphysical hypotheses, and that there is a clear sense in which metaphysical realism is just one, certainly sui generis, metaphysical hypothesis. We thank two anonymous referees for pressing us on this point.

Putting all the above together, it looks like none of the distinctive features of naturalism suffices, either separately or collectively, for concluding that metaphysics and science require the same epistemic attitude in a naturalistic setting. Thus, 4) above is false and, as a consequence, Emery's argument turns out to be unsound.

One may take Emery's side on this and insist that the above entails a difference in goals between science and metaphysics which is unacceptable by naturalistic standards. This, however, would be unmotivated and question-begging. For, given that at no point was it denied that there is an important amount of continuity between science and metaphysics at the level of methodology and ontological commitment, and that metaphysics should never ignore, let alone contradict, science, the above rejection of PT appears perfectly compatible with an overall naturalistic perspective.

Let us illustrate the foregoing with a concise case study. Consider the issue of the nature of quantum objects.[18] Roughly, there are two 'metaphysical packages' available for quantum objects: the objects-qua-individuals and the objects-qua-non-individuals packages, both of which may be cashed out in terms of bundles, tropes, or substances (French & Bigaj, 2024). Which package is the one we find in the real world, and its metaphysical nature, is left entirely underdetermined by scientific evidence. As an alternative, one may cash out quantum entities not in terms of objects but in terms of structures, but metaphysical underdetermination arises there as well (French, 2020); for example, a structure might be metaphysically described in various ways in terms of extrinsic properties, relations, and/or symmetry groups (Bianchi & Giannotti, 2021). Now, crucially, whatever reasons may be adduced to prefer one of these options over the others, they would not be purely empirical. Indeed, for any pair of opposing theses in metaphysics, both metaphysical packages at stake are, as Esfeld (2013, p. 21) puts it, a "[…] a purely metaphysical move that one can always make, physics be as it may". This is because, almost by definition, a metaphysical gloss being added to a scientific theory means that the empirical data that led to the formulation of the theory underdetermines its interpretation.[19] Of course, keeping the empirical facts fixed, one can choose between metaphysical alternatives only on the basis of extra-empirical factors. Since, as we have pointed out, these are plausibly regarded as not truth-conducive, it follows that theory choice in metaphysics is crucially based on non-truth-conducive factors, regardless of the amount of continuity that one manages to establish between one's metaphysical hypotheses and our best current scientific theories. As Ladyman has it, "[…] the scientific and metaphysical underdetermination problems are different, and even in the scientific case it is questionable that inference to the best explanation is needed" (Ladyman, 2012, p. 45). Indeed, in the scientific case, a scientific realist attitude towards quantum mechanics may be regarded as perfectly justified merely on the basis of the extraordinary empirical success of the theory — regardless of whether or not one is in a position to provide a compelling ontological and metaphysical gloss to it.

---

[17] Based on this kind of differences, even those sympathetic to a scientifically-oriented philosophy may claim that — whatever the merits of theoretical virtues in scientific theory choice — they fare systematically worse in metaphysical theory choice. For arguments in this direction, see Ladyman (2012) and Saatsi (2017).

[18] To avoid misunderstanding, what follows is intended merely as an exemplification of the situation that we described in the preceding paragraphs. Our argument is based on the considerations made up to this point, not on the idea that the quantum mechanics example is paradigmatic and can be generalised to metaphysics as a whole.

This latter claim is, of course, likely to be rejected by the 'deep realist' à la French who, as we have seen, believes that metaphysics is part and parcel of one's scientific realism. On the other hand, one may respond that it is in fact the deep realist that owes us an argument in favour of a metaphysically loaded realism; and, lacking that, in view of our earlier considerations the case against metaphysical realism is stronger than the case in favour of the necessity for scientific realism to come equipped with a precise metaphysical picture of the relevant domain.

Be this as it may, it is not our aim here to solve this issue once and for all. Moreover, and more importantly, we believe that there are ways to endow scientific realism with a metaphysical element without ipso facto assuming that one's realism extends automatically to the latter. Getting back to the example used a moment ago, perhaps it may make sense for philosophers to explore issues in the metaphysics of quantum mechanics even if none of the contenders has (and, perhaps, will ever have) sufficient strength to warrant a realist attitude towards it. Let us, then, briefly see what view(s) of (naturalised) metaphysics may support a perspective of this sort, and at the same time bring the present paper to its conclusion.

## 4. Concluding remarks (and a suggestion)

---

[19] We say 'almost by definition' because one may think that metaphysics flows directly from scientific theories — hence, from the underlying empirical input (cfr. scientific metaphysics as discussed in the first section). Although there is no space here to argue for this, we just repeat that this view of the metaphysical implications of science is highly questionable, as witnessed by the degree of disagreement that persists among naturalistic metaphysicians. Of course, there are cases in which scientific theories make certain metaphysical views more natural than others, or even assume such views at the outset. For example, in Bohmian mechanics, space-time individuality is taken for granted as all particles indeed have well-defined space-time positions at all instants (see French & Krause, 2006, § 4.4). There are, however, paradigmatic cases in which no metaphysical theory can be said to be more plausible than all the others just by looking at the relevant scientific theory. This is the case, for instance, for the individuality profile of quantum objects in standard quantum mechanics — which is our example here in this paper. We thank Pablo Acuña for drawing our attention to this point.

If, as we argued, there are no compelling reasons for thinking that in a naturalistic context, scientific realism makes realism about metaphysics highly plausible, then it is possible, and indeed advisable, to abandon PT, i.e., the view that the naturalist ought to have the same epistemic attitude towards science and metaphysics. What follows from this?

First of all, if the foregoing analysis is correct, then naturalism is to be intended not as entailing PT, but rather as requiring that metaphysics (be continuous with science and) not be given a *higher* epistemic status than science. Such status can perfectly be *lower* — just as in the case in which one is a realist about, say, standard quantum mechanics but not about particles conceived of as individual objects (or about any other metaphysics of quantum entities, for that matter). In cases like this, however, one may still contend that there is hope to achieve a better understanding of the relevant domain by developing several, alternative metaphysical explanations of the evidence at hand, while remaining sceptic that any of them cuts nature at its joints in the same way scientific theories do — or at least might.

Thus, the idea that one may consistently endorse scientific realism and a non-eliminative, yet non-realist, attitude towards metaphysics, or be a scientific anti-realist and reject the view that non-naturalism and eliminativism about metaphysics are the only options available to them, is crucially based on a separation of truth, on the one hand, and understanding and explanation on the other.[20]

One good example of such an attitude towards metaphysics is Bueno's (2021) neo-Pyrrhonism. According to neo-Pyrrhonists, metaphysical theorising should be kept at a minimum but, when deemed not eliminable — say, when discussing the (non-)individuality of quantum particles in the context of an explanation of the peculiarities of quantum statistics — it should extend the empiricist way of looking at science to metaphysics as well. Recall that the constructive empiricist says that all the scientific endeavour should aim for is to achieve empirically adequate theories (i.e., not *true* ones). According to Bueno, that conception might be extended to metaphysics. Instead of eliminating the metaphysical enterprise, the neo-Pyrrhonian stance would recommend the following:

> "[…] given the disagreement in question [concerning, say, whether quantum objects are individuals or not], and the fact that the different answers turn out to be all empirically adequate, it is unclear how to choose between the rival conceptions on empirical grounds. One could try to choose, once again, based on methodological considerations. Nevertheless, for the reasons just discussed, it is unclear to what extent this move is likely to succeed. Being unable to decide the issue, neo-Pyrrhonists suspend judgment." (Bueno, 2021, p. 11)

---

[20] One may think that this is a non-starter because understanding and explanation are factive, i.e., require the reality of what explains and/or provides understanding. Instrumentalists clearly deny this, and doing so doesn't look implausible after all, at least as long as potential explanation (and understanding based on it) is concerned.

Now, crucially, suspending judgement about paradigmatically metaphysical questions is in this case different from *eliminating* such questions, as the neo-Pyrrhonian stance aims instead to achieve *understanding* exactly on the basis of those several, strongly underdetermined metaphysical possibilities regarding how the world could be. Although they are mutually exclusive and individually lack the epistemic grounds required for belief, that is, it is nonetheless the case that metaphysical hypotheses are instrumental to achieving very important epistemic aims. McSweeney (2023) seems to go in the same direction, as she argues that metaphysics is an essentially imaginative enterprise aimed not at truth, but rather as improving our understanding of the world around us — even if only on the basis of imagining ways things could be. Emery (2023) and McKenzie (2020) also believe that there is indeed significant actual room in epistemic space between the option of eliminating of questions that appear hard, if not impossible to answer, and the option of believing in (what one takes to be) the one true answer.

We entirely agree with this perspective on metaphysics and metaphysical explanation. Much like science can be deemed highly valuable because it 'saves the phenomena' providing explanation and understanding, but not necessarily truth, based on hypotheses going beyond the observable, so — we claim — metaphysics can be deemed valuable because it furthers our understanding of what is around us, albeit in terms of entities, mechanisms and processes whose existence we may (perhaps forever) be unable to establish beyond doubt, or even just over the threshold for reasonable belief.

Based on this idea that metaphysics (too) may just 'save the phenomena'[21], rather than opting for a suspension of judgment in matters metaphysical, one might find it useful to employ the epistemic notion of acceptance. Crucially, a proposition is accepted when one uses it for certain purposes while refraining from regarding it as true. And this may well be what one does when one takes metaphysical hypotheses to contribute to our *understanding* of the world. Arguably, this may even be made compatible with the traditional conception of metaphysics as in the business of uncovering the fundamental nature of reality. In order to do so, one would have to distinguish between the typical attitude of A) those who find themselves working with an explanatory hypothesis (be it with a view to building it, refining it, modifying it to parry objections or whatever) and B) those reflecting about that hypothesis from the outside, as it were. In this sense, metaphysicians who think hard about ways to make, say, realism about universals more coherent and/or more explanatory would be like experimental physicists in the lab, with a corresponding 'natural ontological attitude' towards the sort of entities they are in the business of inquiring into.

Philosophers thinking about meta-metaphysical issues may instead be more akin to scientists or — perhaps more likely — philosophers of science, pondering more generally about the actual epistemic warrant of scientific claims. This is the sort of approach we favour. It could be understood as a fictionalist attitude towards metaphysics (Rosen 2020; Arroyo & Morganti, 2025) — its distinctive feature being that it does not simply bracket truth altogether, but rather restricts it to 'truth within the fiction' constituted by the specific metaphysical hypothesis one is entertaining.[22] However, going into the details of fictionalism about metaphysics is neither possible nor necessary here. Surely, there are other alternatives to consider — such, as for instance, some sort of neo-Carnapian approach, or some kind of full-blown pragmatism about metaphysics — and the specific way in which non-eliminative, non-realist naturalism about metaphysics is best construed certainly requires further study, which we cannot but leave to future work.

---

[21] Possibly including some scientific unobservables with respect to which realism may be deemed justified.

# Declarations

### Availability of data and material

Data availability statement: not applicable.

### Competing interests

The authors declare that no financial or non-financial interests are directly or indirectly related to the work submitted for publication.

### Funding

Raoni Arroyo was supported by grant #2021/11381-1, São Paulo Research Foundation (FAPESP). The earlier versions of this article were produced during his research visit to the Department of Philosophy, Communication and Performing Arts of the Unviersity of Roma Tre, Rome, Italy, supported by grant #2022/15992-8, São Paulo Research Foundation (FAPESP). Matteo Morganti was supported by the MUR grant PRIN 2022, "The Philosophical Reception of Quantum Theory in France and German-speaking Countries between 1925 and 1945: Conceptual Implications for the Contemporary Debate", prot. 20224HXFLY. He also acknowledges support by the FCT, Portugal, through the grant 2023.15136.PEX "Indeterminacy in Science" (principal investigator: Dr. Robert Michels), https://doi.org/10.54499/2023.15136.PEX.

### Authors' contributions

All authors contributed equally to the conception, development, and writing of this work. All authors approved the final version of the manuscript.

---

[22] It is perhaps worth pointing out explicitly the difference between fictionalism understood in this way, i.e., as a peculiar non-realistic stance towards our claims concerning putatively 'real stuff', and fictionalism as the view that the relevant, in this case, metaphysical, discourse, makes true claims about fictional entities.

**Acknowledgements**

This work was presented at the NEL Work in Progress Seminar (SNEL), Federal University of Santa Catarina (UFSC), Brazil, and at the Workshop sobre Metafísica de las Ciencias, University of Chile. We are especially grateful to Pablo Acuña, Jonas Arenhart, Bruno Borge, Anjan Chakravartty, Ivan da Cunha, Evelyn Erickson, Robert Michels, Joaquim Gianotti, Emanuele Rossanese, and Cristián Soto for their valuable feedback. We also thank two anonymous reviewers for their insightful comments and suggestions.